# High Accuracy Android Malware Detection Using Ensemble Learning

Suleiman Y. Yerima, Sakir Sezer, Igor Muttik

Postprint

# High Accuracy Android Malware Detection Using Ensemble Learning

Suleiman Y. Yerima, Sakir Sezer,
Centre for Secure Information Technologies
Queen's University Belfast,
Belfast, Northern Ireland, United Kingdom
E-mail: {s.yerima, s.sezer, g.mcwilliams}@qub.ac.uk

Igor Muttik
Senior Principal Research Architect
McAfee Labs (Part of Intel Security)
London, United Kingdom
E-mail: igor_muttik@mcafee.com

*Abstract—* With over 50 billion downloads and more than 1.3 million apps in Google's official market, Android has continued to gain popularity amongst smartphone users worldwide. At the same time there has been a rise in malware targeting the platform, with more recent strains employing highly sophisticated detection avoidance techniques. As traditional signature based methods become less potent in detecting unknown malware, alternatives are needed for timely zero-day discovery. Thus this paper proposes an approach that utilizes ensemble learning for Android malware detection. It combines advantages of static analysis with the efficiency and performance of ensemble machine learning to improve Android malware detection accuracy. The machine learning models are built using a large repository of malware samples and benign apps from a leading antivirus vendor. Experimental results and analysis presented shows that the proposed method which uses a large feature space to leverage the power of ensemble learning is capable of 97.3 % to 99% detection accuracy with very low false positive rates.

## 1. INTRODUCTION

There has been a dramatic increase in Android malware since the first SMS Trojan was discovered in the wild by Kaspersky in August 2010. Since then, Android malware have been evolving, becoming more sophisticated in avoiding detection. As the study in [1] revealed, recent malware families exhibit polymorphic behavior, malicious payload encryption, increased code obfuscation, stealthy command and control communications channels, dynamic runtime loading of malicious payload, etc. Not only do these anti-analysis techniques present problems for traditional signature-based detection, but also significantly increase the effort in uncovering malicious behavior and code within the Android applications.

Due to difficulty in spotting malicious behavior, Android malware could remain unnoticed for up to three months on average [2]. Moreover, most antivirus detection capabilities depend on the existence of an updated malware signature repository, therefore the antivirus users are at risk whenever previously un-encountered malware is spread. Since the response time of antivirus vendors may vary between several hours to several days to identify malware, generate a signature, and update their client's signature database, hackers have a substantial window of opportunity [3].

Oberheide et al. also observed that it took on average 48 days for a signature based antivirus engine to become capable of detecting new threats [4]. Although Google introduced Bouncer to its app store in order to screen submitted apps for malicious behavior, it has been shown to be vulnerable to detection avoidance by well-crafted malicious apps [5].

Clearly, there is a need for improved detection approaches given the evolution of Android malware and the urgency to narrow the window of opportunity for the threats posed by emergence of previously unseen strains. Hence, this paper proposes and investigates an effective approach that exploits the merits of static analysis and ensemble machine learning in order to enable zero-day Android malware detection with high accuracy. Different from existing work on Android malware, this paper proposes, develops, and investigates an extensive feature based approach that applies ensemble learning to the problem. The main contributions of the paper are:

• A new high accuracy static analysis driven Android malware detection scheme is proposed, investigated and developed based on ensemble machine learning.
• A large feature space approach is developed employing 179 features for classification decisions. The features are from diverse categories: API calls, commands and permissions, thus inherently resilient to obfuscation within app code as majority of the features will still be accurately extractable. Furthermore, a large malware sample repository is used, which when combined with an extensive features set allows the power of ensemble learning to be fully exploited.
• Presents extensive empirical evaluation studies based on real malware and benign samples from a leading antivirus vendor's repositories giving insights on the efficacy of the proposed scheme.

Our approach is beneficial in several scenarios: from filtering apps to uncover new malicious apps in the wild; prioritizing apps for further (more expensive) analysis; policing app markets; verification of new apps prior to installation, etc. The rest of the paper is organized as follows. In section 2, a survey of related work is presented, followed by the feature extraction process in section 3. Next, ensemble machine learning is discussed in Section 4. Section 5 presents methodology and experiments undertaken while empirical results are presented in section 6. The paper is then concluded in section 7.

## 2. RELATED WORK

Several works such as [6]-[12] apply static analysis for detection of Android malware. Grace et al. proposed RiskRanker [6] for automated risk assessment and app profiling in order to police Android markets. Wei et al. also proposed a profiling tool for Android apps called ProfileDroid [7]. This provides a multi-layer monitoring and profiling system to characterize Android app behaviour in several layers. In [8], Batyuk et al. proposed using static analysis for identifying security and privacy threats. AndroidLeaks [9], SCANDAL [10], and the approach presented in [11] are frameworks that detect privacy information leakage based on static analysis. Furthermore, in [12], the Android Application Sandbox (AAS) is proposed by Blasing et al. AAS uses both static and dynamic analysis, where the static analysis part is based on matching 5 different patterns from decompiled code. Static analysis also provides the basis for the heuristic engine proposed in [2] for detecting Android malware using 39 different flags. A risk score is calculated from the heuristics in order to prioritize the most likely malicious samples for analysis.

Other previous works utilizing static analysis include Comdroid [13] and DroidChecker [14]. Different from [2], and [6]-[14], the approach in this paper leverages automated static analysis for high accuracy and robust learning based malware detection.

Machine learning approaches to malware detection have been previously studied on PC platforms [15]-[19]. Investigation of machine learning based detection for Android platform is gaining attention recently with the growing availability of malware samples. Related work that apply machine learning with static analysis to detect Android malware can be found in [20] - [24] for instance.

In [20], Bayesian classification was applied to categorize apps into 'benign' or 'suspicious' using 58 code-based feature attributes. The training and classification employed 1000 Android malware samples from 49 families and 1000 benign applications. The approach in [21] utilized permissions and call flow graphs for training SVM models to distinguish between benign and malicious Android apps. The authors derived one-class SVM models based on the benign samples alone and use these for identification of both benign and malicious apps. In [22], Sanz et al compared various machine learning schemes trained with permission features on their malware detection accuracy. Their analysis is based on 249 malware samples and 347 benign apps. Sarma et al. [23] and Peng et al. [24] also apply permissions to train SVM based and Bayesian based models respectively for risk ranking of Android apps. D. –J. Wu et al. proposed DroidMat in [25], where Android malware is detected using k-means clustering after computing the required number of clusters by Singular Value Decomposition (SVD) approach. They present experimental results based on 238 Android samples from 34 families together with 1500 benign apps. Our work differs from [20]-[25], as our static analysis driven approach leverages ensemble learning driven by a more extensive feature set comprising 179 feature attributes (from API calls, commands, and permissions). Additionally, our study utilizes a larger malware dataset than the previous works.

In [26], the authors also apply machine learning with static analysis, but utilize Linux malware rather than Android malware samples. Their approach extracts Linux system commands within Android and use the readelf command to output a list of referenced function calls for each system command. Some proposed Android malware detection methods are based on dynamic analysis. Shabtai et al. [27], [28] proposed a host-based solution that continuously monitors various features and events like CPU consumption, number of packets sent, number of running processes, keyboard/touch-screen pressing etc. Machine learning anomaly detectors are then applied to classify the collected data into normal or abnormal. M. Zhao et al. propose AntiMalDroid in [29]. AntiMalDroid is a software behavior signature based malware detection framework that uses SVM to detect malware and their variants in runtime and extend malware characteristics database dynamically. Logged behaviour sequence is used as the feature for model training and detection. Crowdroid [30] is proposed as a behavior-based malware detection system that traces system calls behavior, converts them into feature vectors and applies k-means algorithm to detect malware. Enck et al. [31] perform dynamic taint tracking of data in order to reveal to a user when an app is trying to exfiltrate sensitive data.

## 3. APPLICATION FEATURE EXTRACTION

In order to obtain the features used in the machine learning based detection, an extended version of our Java based APK analysis tool described in [20] was used. As shown in Figure 1, the tool is enhanced with database storage of feature vectors extracted from app corpus to drive the training phase. The tool, which embeds a Baksmali disassembler [32], can also classify unknown apps using trained models.

The feature extraction is accomplished using feature detectors that extract 65 features comprising critical (Android and Java) API calls and commands as described in [20]. These make up the 'a*pplications attributes*' feature set. A further 130 features are extracted using permissions detector that mines the 'Manifest file' to detect the app permissions.

### 3.1 Proposed feature set

For the purpose of our research, we developed 65 features (54 of which were subsequently used) from API calls and (Linux/Android) command sets. The APIs include (SMS manager APIs (for sending, receiving, reading SMS messages etc.); Telephony manager APIs (for accessing device ID, subscriber ID, network operator, SIM serial number, line number etc.); Package manager APIs (for listing installed packages, installing additional packages etc.).

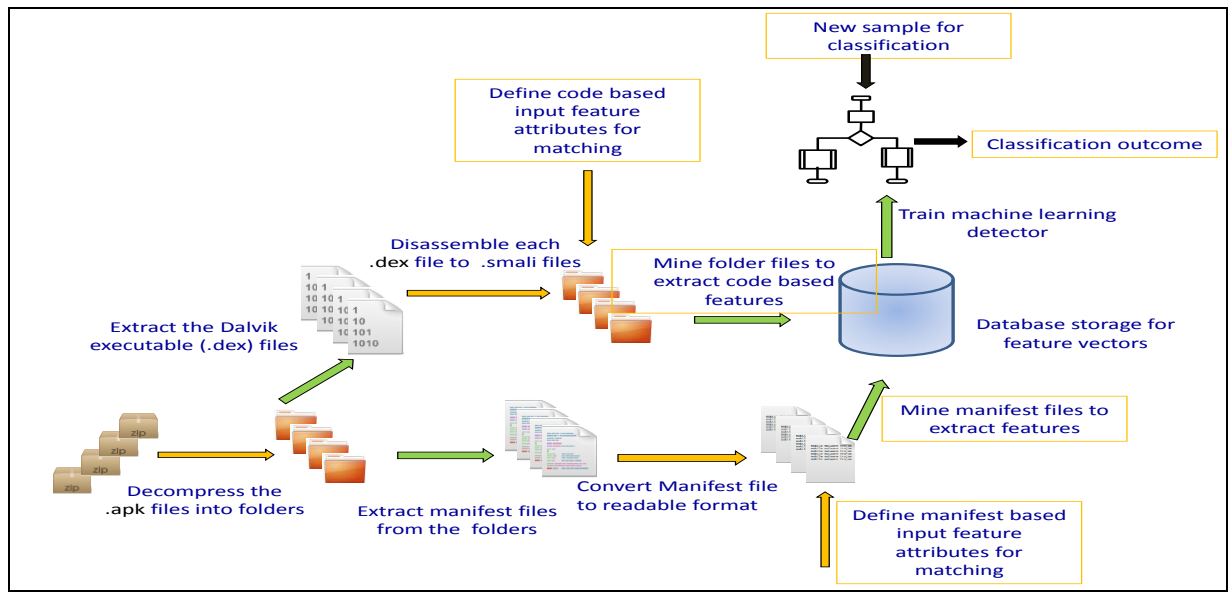

Figure 1: Custom built APK analyzer for feature extraction and classification to identify malicious Android applications.

The API calls feature extraction also includes detection (within the disassembled .smali code) of Java API calls for encryption, reflection, JNI usage, dynamic class loading, creation of new processes, and runtime execution of processes.

Features were also derived from specific Linux commands e.g. shell commands like 'chmod', 'chown', 'mount', and specific strings like '/sys/bin/sh' etc. which enable malware to escalate privilege, root devices, or execute malicious shell scripts at run time. Other type of commands defined as features include Android commands like 'pm install' which can be used for stealthy installation of additional malicious packages. All APK subfolders are also inspected for extracting command based features. Permissions extracted from the Manifest file provided additional set of features. Although all of the 130 standard permissions were included, only 125 were subsequently used (see section 5). An extensive and diverse set of features were chosen for the following reasons:

a) **Robustness**: each set of features are extracted from different parts of the APK (API calls from dex executable, permissions from the Manifest file, and commands are mostly detected outside of the app's executable). Should feature extraction from dex executable fail for example, the permissions would still be obtainable from the Manifest file. Furthermore, if for instance, malware incorporates encryption to prevent detection of commands, API call features (including the crypto API) will still be detectable along with permissions to expose the app's malicious intent. This measure of resilience is enabled by the fact that *all of the features will be utilized in the ensemble learning approach* unlike with other algorithms e.g. Naïve Bayes were feature ranking and selection are necessary for optimum performance. Our approach is contrary to most existing work where ranking and selection steps are used to reduce the feature set. In that regard, such existing approaches will have a less resilient feature set compared to ours.

b) **Performance**: the diversity and extent of the features employed is actually advantageous to ensemble learning as it provides greater degree of freedom to introduce randomness in feature selection. Additionally, by employing a large malware repository in our work, both the *features* and *instances* provide randomization opportunities which enhance ensemble learning performance as explained further in section 4.

4. ENSEMBLE MACHINE LEARNING

Ensemble learning builds a prediction model by combining the strengths of a collection of simpler base models [33]. For our classification problem the application of Random Forest, an ensemble learning method, is proposed to learn the app characteristics based *on all of* the 179 features (*unlike previous machine learning-based approaches that advocate a pre-training stage of 'ranking and feature reduction' for improved performance*). The model can then be used in classifying new Android applications into suspicious or benign. Random Forest combines random Decision Trees with *Bagging* to achieve very high classification accuracy [34]. This paper focuses on investigating how the power of ensemble learning can be applied to improve Android malware detection. By means of datasets composed of extracted features from a large repository of malware and benign apps, Random Forest based classification is investigated under different experimental scenarios. Furthermore, comparative analysis is made to Naïve Bayes, Decision Trees, Random Trees, and Simple Logistic (another ensemble learning technique based on *boosting*).

Decision Trees (DT) are sequential models, which logically combine a sequence of tests that compare a numeric attribute against a threshold value or a nominal attribute against a set of possible values [35]. DT algorithms select the next best feature during splits by employing *information entropy* or *Gini impurity* respectively given by (1) and (2):

$$I_E(f) = -\sum_{i=1}^{m} f_i \log_2 f_i \quad (1)$$

$$GI(f) = \sum_{i=1}^{m} f_i(1-f_i) = \sum_{i=1}^{m} f_i - f_i^2 = 1 - \sum_{i=1}^{m} f_i^2 \quad (2)$$

where $f_i$ is the fraction of items labelled with value $i$ from $m$ categories and GI is known as the *Gini Index*.

The Random Tree algorithm [36] departs from the tradition DT method by testing a given number of random features at each node of the tree and performs no pruning. Random Forest uses *Bagging* (bootstrap aggregation) to produce a diverse ensemble of Random Trees independently trained on distinct bootstrap samples obtained by random sampling the given set N' <= N times with replacement.

*4.1 Random forest algorithm*

Random Forest applies *Bagging* to generate a diverse ensemble of classifiers by introducing randomness into the learning algorithms input [37]. Diversity is also achieved by random feature subset selection during node splitting [34]. Hence, our classification scenario where we apply all 179 features *could benefit from these two dimensions of randomness to improve accuracy*. The Random Forest algorithm is summarized in Figure 2. The training variables are the number of trees *T* and the number of features *m* from the input feature space to be randomly selected at each split of the base tree construction.

Random Forest has several advantages that can be leveraged for improved machine learning based detection: *no special preprocessing of input is required; can deal with large numbers of training instances, missing values, irrelevant features*, etc. More importantly training and prediction phases are both fast, and they are more amenable to parallelization than *Boosting*-based ensemble learners (e.g. Simple Logistic).

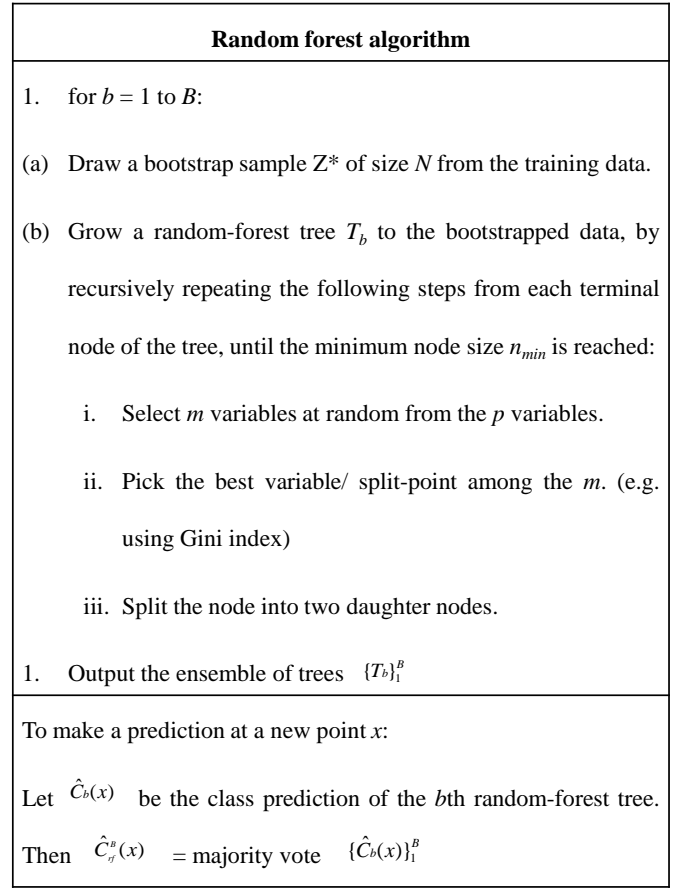

| Random forest algorithm |
|---|
| 1. for *b* = 1 to *B*: <br><br> (a) Draw a bootstrap sample Z* of size *N* from the training data. <br><br> (b) Grow a random-forest tree $T_b$ to the bootstrapped data, by recursively repeating the following steps from each terminal node of the tree, until the minimum node size $n_{min}$ is reached: <br><br>     i. Select *m* variables at random from the *p* variables. <br><br>     ii. Pick the best variable/ split-point among the *m*. (e.g. using Gini index) <br><br>     iii. Split the node into two daughter nodes. <br><br> 1. Output the ensemble of trees $\{T_b\}_1^B$ |
| To make a prediction at a new point *x*: <br><br> Let $\hat{C}_b(x)$ be the class prediction of the *b*th random-forest tree. <br><br> Then $\hat{C}_{rf}^B(x)$ = majority vote $\{\hat{C}_b(x)\}_1^B$ |

Figure 2: The Random Forest algorithm.

*4.2 Simple logistic algorithm*

Simple Logistic is an ensemble learning method based on *'Boosting'*. Simple Logistic utilizes *additive logistic regression* using simple regression functions as base learners. The base learner in Simple Logistic is a regression model based on a single attribute (feature), i.e. the one that maximizes the log-likelihood:

$$L = \sum_{i=1}^{n} \begin{array}{l} (1-x^{(i)})\log(1-\Pr[1|f_1^{(i)}, f_2^{(i)}, ..., f_m^{(i)}]) \\ + x^{(i)} \log(\Pr[1|f_1^{(i)}, f_2^{(i)}, ..., f_m^{(i)}]) \end{array} \quad (3)$$

where *m* is the number of features.

Logistic regression models are built and fitted using *LogitBoost* [37] which performs additive logistic regression. Cross-validation is used to determine the number of iterations to perform which also supports automatic attribute/feature selection [38] [39]. The additive logistic regression algorithm is summarized in Figure 3.

| Additive logistic regression |
|---|
| Model generation |
| **for** $j = 1$ to $t$ iterations **do**: |
|    **for** each feature vector $\mathbf{f}[i]$ **do**: |
|       1: set the target value for the regression to |
|          $c[i] = (y[i] - P(1|\mathbf{f}[i]))/[P(1|\mathbf{f}[i]) * (1 - P(1|\mathbf{f}[i]))]$ |
|       2: set the weight $w[i]$ of instance $\mathbf{f}[i]$ to : |
|          $P(1|\mathbf{f}[i]) * (1-P(1|\mathbf{f}[i]))$ |
|       3: fit a regression model $r[j]$ to the data with |
|          class values $c[i]$ and weights $w[i]$ |
| Classification |
| Predict class 1 if $P(1|\mathbf{f}) > 0.5$, otherwise predict class 0 |

Figure 3: Algorithm for binary additive logistic regression.

## 5. METHODOLOGY AND EXPERIMENTS

The custom built APK analyzer shown in Figure 1 was utilized in extracting the selected features from a collection of benign apps and recent malware samples. The analyzer was run across a total of 6,863 applications (obtained from McAfee's internal repository); 2925 of these were malware while 3938 apps were benign. The extracted features were converted into binary feature vectors with a 0 or 1 indicating the absence/presence of a feature. These were stored in a MySQL database for the training and testing phases.

Out of the initial 65 non-permission based features 54 of these were selected to produce feature vectors for the training phase. The eliminated features were those with no occurrences in either class. Out of the 130 permissions features, 5 had no occurrence in either class i.e.: ADD_VOICEMAIL, SET_POINTER_SPEED, USE_SIP, WRITE_PROFILE, WRITE_SOCIAL_STREAM. Hence the experiments were based on the remaining 125 permission features and 54 API and command based features yielding a total of 179 training features.

In order to investigate the effect of feature diversity, three separate feature sets were created for training models and comparative analysis: (a) Feature set consisting of vectors from the 54 application attribute features only (b) Feature set of vectors from the 125 permissions only (c) Feature set consisting of vectors from a mix of all the (diverse) 179 property vectors (Table 1).

Table 1: Feature sets for model building

| Feature set | Number of features |
|---|---|
| App attributes features (AF) | 54 |
| Permission features (PF) | 125 |
| Combined app attributes and Permission features (CAPF) | 179 |

The learning algorithms were investigated with the three feature sets in order to evaluate their classification performance using the following evaluation metrics. Accuracy and Error Rate are respectively given by:

$$Acc = \frac{n_{ben \to ben} + n_{sus \to sus}}{n_{ben \to ben} + n_{ben \to sus} + n_{sus \to ben} + n_{sus \to sus}} \quad (4)$$

$$Err = \frac{n_{ben \to sus} + n_{sus \to ben}}{n_{ben \to ben} + n_{ben \to sus} + n_{sus \to ben} + n_{sus \to sus}} \quad (5)$$

The false positive rate (FPR), false negative rate (FNR), true positive rate (TPR), true negative rate (TNR) and precision (p) are defined as follows:

$$FPR = \frac{n_{ben \to sus}}{n_{ben \to sus} + n_{ben \to ben}} \quad (6)$$

$$FNR = \frac{n_{sus \to ben}}{n_{sus \to sus} + n_{sus \to ben}} \quad (7)$$

$$TPR = \frac{n_{sus \to sus}}{n_{sus \to ben} + n_{sus \to sus}} \quad (8)$$

$$TNR = \frac{n_{ben \to ben}}{n_{ben \to sus} + n_{ben \to ben}} \quad (9)$$

$$p = \frac{n_{sus \to sus}}{n_{ben \to sus} + n_{sus \to sus}} \quad (10)$$

## 6. RESULTS AND DISCUSSIONS

Results of the experiments undertaken are discussed in this section. Note that all results are obtained using 10-fold cross validation where training and testing sets are different.

*6.1 Naïve Bayes results*

Results from the Naïve Bayes algorithm are presented in Table 2 showing performance of PF (permissions feature set with 125 features), AF (application attributes feature set with 54 features) and several configurations of the CAPF (combined feature sets with 179 features): i.e. top 10, 15, 20, and 25 features ranked using Mutual Information feature selection [40].

With Naïve Bayes, the best detection rate of 85.4% was obtained with the top 10 features from the CAPF set. This is depicted in the Bayes (10) CAPF row of Table 2. The top 20 mixed features from MI ranking together with their frequency of appearance over the entire 6, 863 labeled app instances are shown in Table 3. They include SEND_SMS, RECEIVE_SMS, READ_SMS which are from the permissions feature set, and also '*remount*', '*/system/app*', '*chown*', '*createSubprocess*', which are from the app attributes feature set. The results in Table 2 shows that with the Naïve Bayes algorithm, for best detection rates, the CAPF feature set should be utilized. Note that all the results of the experiments were obtained using 10-fold cross validation.

*6.2 Simple Logistic, Decision Tree and Random Tree results*

Simple Logistic results are presented in Table 4, while those of Decision Tree and Random Trees are given in Tables 5 and 6 respectively. In terms of comparative performance with corresponding feature sets, Simple Logistic performed better than Bayesian. When trained with the 179 mixed features (CAPF), a detection rate of approximately 91% is achieved with 4.6% false positive rate. With Decision Tree classification, TPR of 94.8% and corresponding FPR of 4% were obtained using the CAPF set. As with the Simple Logistics and Naïve Bayes, the CAPF Decision Tree classifier enables better detection rates than either PF or AF based Decision Tree classifiers. On the other hand, the false positive rates for CAPF and AF feature sets were similar, but also better than that of PF. Overall AUC performance is good for both AF feature set and CAPF feature set Decision Trees. Therefore, an FPR-TPR trade off to improve detection rates is feasible (given the low FPRs of 3.9 % and 4 % respectively).

Table 2: Results from Naïve Bayes classifiers

| Feature set | TPR | TNR | FPR | FNR | ACC | ERR | AUC |
|---|---|---|---|---|---|---|---|
| Bayes (179) CAPF | 0.821 | 0.913 | 0.087 | 0.179 | 0.867 | 0.133 | 0.915 |
| Bayes (25) CAPF | 0.826 | 0.918 | 0.082 | 0.174 | 0.872 | 0.128 | 0.906 |
| Bayes (20) CAPF | 0.844 | 0.921 | 0.079 | 0.156 | 0.883 | 0.118 | 0.908 |
| Bayes(15) CAPF | 0.829 | 0.918 | 0.082 | 0.171 | 0.874 | 0.127 | 0.903 |
| Bayes (10) CAPF | **0.854** | 0.913 | 0.087 | 0.146 | 0.884 | 0.117 | 0.895 |
| Bayes (54) AF | 0.565 | 0.928 | 0.072 | 0.435 | 0.747 | 0.253 | 0.873 |
| Bayes (125) PF | 0.68 | 0.886 | 0.114 | 0.32 | 0.783 | 0.217 | 0.86 |

Table 3: Top 20 CAPF features ranked using Mutual Information.

| Feature | Benign | Malware | Information gain score |
|---|---|---|---|
| SEND_SMS | 128 | 1557 | 0.260525 |
| RECEIVE_SMS | 127 | 976 | 0.126554 |
| READ_SMS | 140 | 900 | 0.107046 |
| remount | 30 | 628 | 0.098938 |
| /system/app | 55 | 687 | 0.098179 |
| chown | 51 | 668 | 0.096293 |
| createSubprocess | 5 | 531 | 0.096111 |
| WRITE_SMS | 89 | 720 | 0.090689 |
| /system/bin/sh | 36 | 596 | 0.089475 |
| mount | 146 | 810 | 0.088369 |
| abortBroadcast | 48 | 618 | 0.08799 |
| READ_PHONE_STATE | 2016 | 2378 | 0.072633 |
| TelephonyManager | 2168 | 2451 | 0.069811 |
| TelephonyManager_getSubscriberId | 480 | 1094 | 0.063550 |
| chmod | 459 | 999 | 0.053325 |
| Ljava_net_URLDecoder | 1539 | 445 | 0.051456 |
| ACCESS_NETWORK_STATE | 2973 | 1453 | 0.051394 |
| RESTART_PACKAGES | 142 | 597 | 0.050407 |
| CHANGE_WIFI_STATE | 297 | 756 | 0.048716 |
| Ljavax_crypto_spec_SecretKeySpec | 1719 | 592 | 0.044834 |

Table 4: Results from Simple logistic classifiers

| Feature set | TPR | TNR | FPR | FNR | ACC | ERR | Prec. | AUC |
|---|---|---|---|---|---|---|---|---|
| PF | 0.801 | 0.924 | 0.076 | 0.199 | 0.863 | 0.138 | 0.886 | 0.925 |
| AF | 0.835 | 0.943 | 0.057 | 0.165 | 0.889 | 0.111 | 0.916 | 0.938 |
| CAPF | 0.909 | 0.954 | 0.046 | 0.091 | 0.932 | 0.069 | 0.937 | 0.977 |

Table 5: Results from Decision tree classifiers

| Feature set | TPR | TNR | FPR | FNR | ACC | ERR | Prec. | AUC |
|---|---|---|---|---|---|---|---|---|
| PF | 0.87 | 0.938 | 0.062 | 0.13 | 0.904 | 0.096 | 0.912 | 0.934 |
| AF | 0.939 | 0.961 | 0.039 | 0.061 | 0.950 | 0.050 | 0.948 | 0.967 |
| CAPF | 0.948 | 0.960 | 0.04 | 0.052 | 0.954 | 0.046 | 0.946 | 0.964 |

The results from Random Tree classifier under different parameter settings are shown in Table 6. The number of random variables selection at each split during tree construction (i.e. $k$), is varied from $\log_2 f+1$, to 20 and 50 respectively. For each $k$ configuration, the PF, AF and CAPF feature sets were applied to train and evaluate the Random Tree classifiers.

Table 6: Results from Random tree classifiers

| Feature set (k=log2f+1) | TPR | TNR | FPR | FNR | ACC | ERR | Prec. | AUC |
|---|---|---|---|---|---|---|---|---|
| PF | 0.901 | 0.928 | 0.072 | 0.099 | 0.915 | 0.085 | 0.903 | 0.934 |
| AF | 0.946 | 0.949 | 0.051 | 0.054 | 0.948 | 0.053 | 0.933 | 0.952 |
| CAPF | 0.955 | 0.952 | 0.048 | 0.045 | 0.954 | 0.047 | 0.936 | 0.954 |
| Feature set K=20 | TPR | TNR | FPR | FNR | ACC | ERR | Prec. | AUC |
| PF | 0.902 | 0.922 | 0.078 | 0.098 | 0.912 | 0.088 | 0.895 | 0.933 |
| AF | 0.948 | 0.95 | 0.05 | 0.052 | 0.949 | 0.051 | 0.934 | 0.955 |
| CAPF | 0.96 | 0.956 | 0.044 | 0.04 | 0.958 | 0.042 | 0.942 | 0.960 |
| Feature set K=50 | TPR | TNR | FPR | FNR | ACC | ERR | Prec. | AUC |
| PF | 0.898 | 0.928 | 0.072 | 0.102 | 0.913 | 0.087 | 0.902 | 0.933 |
| AF | 0.951 | 0.951 | 0.049 | 0.049 | 0.951 | 0.049 | 0.935 | 0.958 |
| CAPF | 0.961 | 0.956 | 0.044 | 0.039 | 0.959 | 0.041 | 0.942 | 0.960 |

*6.3 Random Forest results*

For the Random forest algorithms, two parameters needed for training the models are the number of trees in the ensemble, T, and the number of random variables selection at each split during tree construction, k. As with the Random Tree classifier, k is varied from log2f +1, to 20 and 50 respectively; and for each k configuration, the three feature sets were compared. The results for T = 10 trees (from 10-fold cross-validation) are given in Table 7.

Table 7: Results from Random forest classifiers.

| Feature set (k=log2f+1) | TPR | TNR | FPR | FNR | ACC | ERR | Prec. | AUC |
|---|---|---|---|---|---|---|---|---|
| PF | 0.901 | 0.949 | 0.051 | 0.099 | 0.925 | 0.075 | 0.929 | 0.966 |
| AF | 0.954 | 0.971 | 0.029 | 0.046 | 0.963 | 0.037 | 0.961 | 0.987 |
| CAPF | 0.971 | 0.977 | 0.023 | 0.029 | 0.974 | 0.026 | 0.969 | 0.992 |
| Feature set K=20 | TPR | TNR | FPR | FNR | ACC | ERR | Prec. | AUC |
| PF | 0.898 | 0.944 | 0.056 | 0.102 | 0.921 | 0.079 | 0.922 | 0.969 |
| AF | 0.956 | 0.969 | 0.031 | 0.044 | 0.963 | 0.037 | 0.958 | 0.987 |
| CAPF | **0.972** | **0.975** | **0.025** | **0.028** | **0.974** | **0.026** | **0.967** | **0.993** |
| Feature set K=50 | TPR | TNR | FPR | FNR | ACC | ERR | Prec. | AUC |
| PF | 0.901 | 0.94 | 0.06 | 0.099 | 0.921 | 0.079 | 0.918 | 0.966 |
| AF | 0.955 | 0.961 | 0.039 | 0.045 | 0.958 | 0.042 | 0.948 | 0.986 |
| CAPF | **0.973** | **0.977** | **0.023** | **0.027** | **0.975** | **0.025** | **0.969** | **0.993** |

As with the Random Tree, the results of the Random Forest classification are not very sensitive to variation in $k$. However, the best detection rates, lowest false positive rates and largest AUC occur simultaneously with the CAPF feature set. *This not only confirms the robustness of Random Forest to large number of input training features, but also justifies our large (179) feature set approach which yields high fidelity malware detection via improved accuracy.* Also, the number of trees T had a negligible impact on performance when increased from 10 to 50 (for all the different values of $k$) and are therefore omitted.

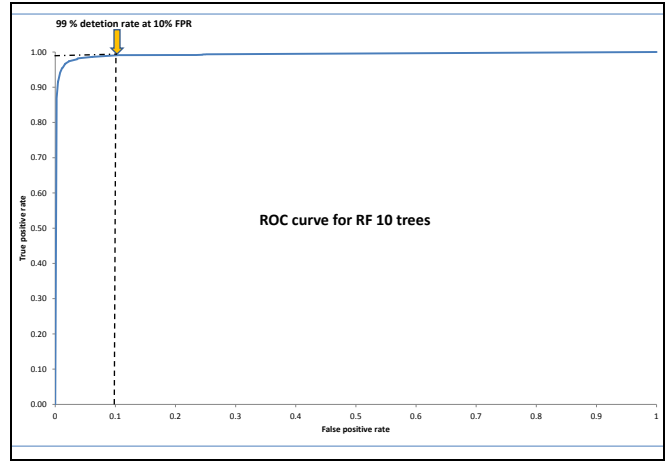

Figure 4: ROC curve for Random Forest.

From Table 7, a very high classification accuracy of 97.6% is observed in the $k$ = 50 configuration. Malware detection rates (TPR) are 97.2% and 97.3 % with 2.5% and 2.3% false positive rates for $k$=20 and $k$=50 respectively. These results outperform all the previously discussed algorithms. Also, Figure 4 illustrates the AUC for the Random Forest classifier of mixed features for $k$= 50. It can be seen that with ROC area of 0.993, malware detection rate can be improved to **98.6%** for **6.3%** false positive rate and **99%** for **10%** false positive rate. These higher TPR operating points will suffice for some application scenarios e.g. filtering apps to prioritize resource allocation for further/manual analysis.

Figures 5 and 6 show graphical comparisons of the different classifier performances (with CAPF features set). Not only does the Random Forest learner perform very well with our mixed features dataset, but also model building time was quite fast. Random Forest (10 trees) learning with $k= 8$, 20 and 50 respectively take 1.48s, 2.96s and 6.41 seconds to build. With Decision Tree classifiers, the requirement for pruning increases the model build time over Random Trees. Simple Logistic learning took longest to build in 81.9 seconds. This is due to the additive nature of the underlying Boosting algorithm which incrementally builds the model based on previous base learners.

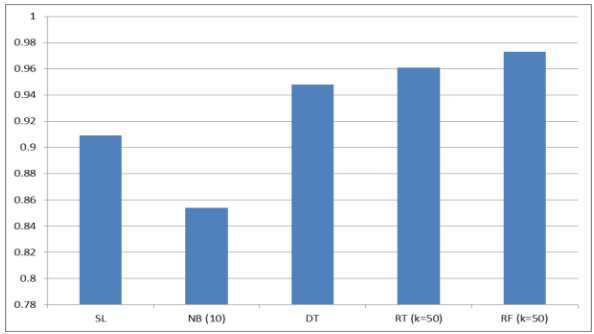

Figure 5: Detection rates for different classifiers.

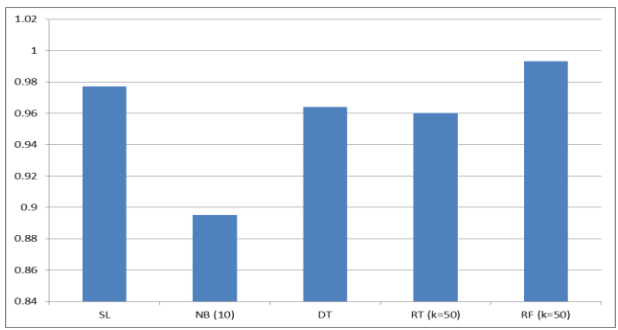

Figure 6: Area under ROC curve for different classifiers.

*6.4 Results comparison with existing work*

In order to highlight the significance of our results, a comparison is made with those published in other static analysis based works where a quantitative comparison is possible using similar metrics. Table 8 shows how the results in this paper measures against the best results of [1], [20], [22], [23], [24], [33], and [41] respectively (using the available metrics). These comparative results show that our approach in this paper outperforms previous similar efforts. This is highlighted in the bottom of the table where the result of our proposed approach that leverages the power of ensemble learning clearly shows the highest detection accuracy and the best AUC performance.

Table 8: Result comparison with existing work.

| Related work | No of samples Mal/ben | TPR | TNR | FPR | ACC | ERR | AUC |
|---|---|---|---|---|---|---|---|
| Yerima et al. [20] | 1000/1000 | 0.906 | 0.932 | 0.068 | 0.918 | 0.082 | 0.974 |
| PUMA [22] | 249/357 | 0.92 | | 0.21 | 0.858 | | 0.920 |
| Kirin [33] | - | 0.39 | | 0.04 | | | |
| Zhou et al. [1] | 1260/0 | 0.769 | - | - | - | - | - |
| DroidMat [25] | 238/1500 | 0.873 | - | - | 0.9787 | - | - |
| Peng et. al [24] | 378/- | - | - | - | - | - | 0.94 - 0.96 |
| Sarma et. al [23] | 121/158,062 | - | - | - | - | - | 0.85-0.91 |
| Yerima et. al. [41] | 1000/1000 | 0.909 | 0.949 | 0.051 | 0.931 | 0.069 | 0.977 |
| **CAPF with RF (T=10, k=50)** | 2925/3938 | **0.973** | **0.977** | **0.023** | **0.975** | **0.025** | **0.993** |

## 7. CONCLUSION

This paper presented a new ensemble learning based Android malware detection approach which can effectively improve detection rates to 97-99% with low false positives by harnessing large mixed feature sets in ways infeasible with traditional machine learning. With this approach there is no requirement for feature selection step to eliminate 'less relevant' features. This use of an extensive mixed feature set provides robustness and resilience to code obfuscation and other anti-analysis techniques being employed by malware authors vastly improving the chances of prompt zero-day malware detection. Experiments performed with large malware dataset from a leading AV vendor demonstrate the effectiveness of the proposed scheme and the higher fidelity achievable compared to traditional approaches.

## REFERENCES

[1] Zhou, Y., and Jiang, X.: 'Dissecting android malware: Characterization and evolution'. Proc. IEEE Symp. on Security and Privacy (SP), San Fransisco, CA, USA, May 2012, pp. 95-109

[2] Apvrille, A., and Strazzere, T.: 'Reducing the window of opportunity for Android malware Gotta catch 'em all', Journal in Computer Virology, 2012, Vol. 8, No. 1-2, pp. 61-71

[3] Dagon, C., Martin, T., and Starner, T.: 'Mobile phones as computing devices the viruses are coming', Pervasive computing, 3, pp. 11-15

[4] Oberheide, J., Cooke, E., and Jahanian, F.: 'Cloudav: N-version antivirus in the network cloud'. Proc. 17th USENIX Security Symposium (Security '08), San Jose, CA, July 2008, pp. 91-106

[5] Oberheide, J., and Miller, C.: 'Dissecting the Android Bouncer'. SummerCon 2012, Brooklyn, NY, USA, June 2012

[6] Grace, M., Zhou, Y., Zhang, Q., Zou, S., and Jiang. X.: 'RiskRanker: scalable and accurate zero-day android malware detection'. Proc. 10th Int. Conf. on Mobile Systems, Applications, and Services (MobiSys '12) ACM, Lake District, UK, June 2012, pp. 281-294

[7] Wei, X., Gomez, L., Neamtiu, I., and Faloutsos. M.: 'ProfileDroid: multi-layer profiling of android applications' Proc. 18th Int. Conf. on

Mobile Computing and Networking (Mobicom '12). Istanbul, Turkey, August 2012, pp. 137-148

[8] Batyuk, L., Herpich, M., Camtepe, S. A., Raddatz, K., Schmidt, A., Albayrak, S.: 'Using static analysis for automatic assessment and mitigation of unwanted and malicious activities within android applications'. 6th Int. Conf. on Malicious and Unwanted Software (MALWARE 2011), Fajardo, PR, USA, October 2011, pp. 66-72

[9] Gibler, C., Crussell, J., Erickson, J., and Chen, H.: 'AndroidLeaks: automatically detecting potential privacy leaks in android applications on a large scale'. Proc. 5th Int. Conf. on Trust and Trustworthy Computing (TRUST 2012), Vienna, Austria, June 2012, pp. 291-307

[10] Kim, J., Yoon, Y., Yi, K., and Shin, J.: 'SCANDAL: static analyzer for detecting privacy leaks in android applications'. Mobile Security Technologies, MoST 2012, San Francisco, May 2012

[11] Mann, C., and Starostin, A.: 'A framework for static detection of privacy leaks in android applications'. Proc. 27th Annual ACM Symposium on Applied Computing (SAC '12), Trento, Italy, March 2012, pp. 1457-1462

[12] Bläsing, T., Batyuk, L., Schmidt, A.-D., Camtepe, S. A., Albayrak S.: 'An android application sandbox system for suspicious software detection'. 5th Int. Conf. on Malicious and Unwanted Software (MALWARE 2010), Nancy, France, Oct. 2010, pp. 55-62

[13] Chin, E., Felt, A. P., Greenwood, K., and Wagner, D.: ' Analyzing inter-application communication in android'. Proc. 9th Int. Conf. on Mobile Systems, Applications, and Services (MobiSys '11). ACM, Washington, DC, USA, June 2011, pp. 239-252

[14] Chan, P. P.F., Hui, L. C.K., and Yiu. S. M.: 'DroidChecker: analyzing android applications for capability leak'. Proc. fifth ACM Conf. on Security and Privacy in Wirelessand Mobile Networks (WISEC '12), Tucson, AZ, USA, April 2012, pp. 125-136

[15] Schultz, M. G., Eskin, E., Zadok, E., and Stolfo, S. J.: 'Data mining methods for detection of new malicious executables'. Proc. 2001 IEEE Symposium on Security and Privacy, (SP '01), Oakland, CA, USA, May 2001, pp. 38-49

[16] Wang, T-Y., Wu, C-H., Hsieh, C-C.: 'A virus prevention model based on static analysis and data mining methods'. Proc. IEEE 8th Int. Conf. on Computer and Information Technology Workshops, Sydney, July 2008, pp. 288-293

[17] Chen, Y., Narayanan, A., Pang, S., and Tao, B.: 'Malicious software detection using multiple sequence alignment and data mining'. 26th IEEE Int. Conf. on Advanced Information Networking and Applications AINA 2012

[18] Santos, I., Brezo, F., Sanz B., Laorden, C., Bringas. P.G.: 'Using opcode sequences in single-class learning to detect unknown malware', IET inf. Secur., 2011, Vol. 5, Iss. 4, pp. 220-227

[19] Muttik, I.: 'Malware mining'. Proc. 21st Virus Bulletin International Conference, VB2011, 5-7 Oct. 2011, Barcelona,Spain.

[20] Yerima, S. Y., Sezer, S., McWilliams, G., Muttik, I.: 'A new android malware detection approach using bayesian classification'. Proc. 27th IEEE int. Conf. on Advanced Information Networking and Applications (AINA 2013), Barcelona, Spain.

[21] Sahs J., and Khan L.: 'A Machine Learning Approach to Android Malware Detection'. Proc. of European Intelligence and Security Informatics Conference, Odense, Denmark, August 2012, pp. 141-147

[22] Sanz, B., Santos, I., Laorden, C., Ugarte-Pedro, X., Bringas, P. G., Alvarez G.: 'PUMA: Permission Usage to Detect Malware in Android'. International Joint Conference CISIS'12-ICEUTE´12-SOCO´12 Special Sessions, in Advances in Intelligent Systems and Computing, Volume 189, pp. 289-298

[23] Sarma, B., Gates, C., Li, N., Potharaju, R., Nita-Rotaru, C., Molloy. I.: 'Android permissions: A perspective combining risks and benefits'. Proc. 17th ACM Symposium on Access Control Models and Technologies, (SACMAT '12), June 2012, pp. 13-22

[24] Peng, H., Gates, C., Sarma, B., Li, N., Qi, A., Potharaju, R., Nita-Rotaru, C., and Molloy, I.: 'Using probabilistic generative models for ranking risks of Android apps'. Proc. of the 19th ACM Conference on Computer and Communications Security (CCS 2012), Raleigh, NC, USA, Oct. 2012, pp. 241-252

[25] Dong-Jie, W., Ching-Hao, M., Te-En, W., Hahn-Ming, L., and Kuo-Ping, W.: 'DroidMat: Android malware detection through manifest and API calls tracing', Proc. Seventh Asia Joint Conference on Information Security (Asia JCIS), 2012, pp. 62-69

[26] Schmidt, A.-D., Bye, R., Schmidt, H.-G., Clausen, J., Kiraz, O., Yuksel, K. A., Camtepe, S. A., Albayrak, S.: 'Static analysis of executables for collaborative malware detection on Android'. IEEE International Conference on Communications, (ICC '09), Dresden, Germany, June 2009, pp.1-5

[27] Shabtai, A., and Elovici, Y.: 'Applying behavioral detection on android based devices'. In MOBILWARE, 2010 pp 235–249

[28] Shabtai, A., Kanonov, U., Elovici, Y., Glezer, C., and Weiss, Y.: 'Andromaly: a behavioral malware detection framework for android devices'. J. Intell. Inf. Syst., 2012, Vol. 38 (1), pp. 161–190

[29] Zhao, M., Ge, F., Zhang, T., and Yuan, Z.: Antimaldroid: 'An efficient svm based malware detection framework for android'. In Communications in Computer and Information Science, Springer, 2011, Vol. 243, pp. 158–166

[30] Burguera, I., Zurutuza, U., and Nadjm-Tehrani, S.: 'Crowdroid: behavior-based malware detection system for Android'. Proc. 1st ACM Workshop on Security and Privacy in Smartphones and Mobile devices (SPSM '11), New York, USA, 2011, pp. 15–26

[31] Enck, W., Gilbert, P., Chun, B., Cox, L., Jung, J., McDaniel, P., and Sheth., A.: 'Taintdroid: An information-flow tracking system for realtime privacy monitoring on smartphones'. In Proc. 9th USENIX conference on Operating systems design and implementation, USENIX 2010, pp. 1-6

[32] http://code.google.com/p/smali, Accessed October 2014

[33] Hastie, T., Tibshirani, R., Freidman, J.: 'The Elements of Statistical Learning' (Springer, 2001)

[34] Breiman, L.: 'Random forests'. Machine Learning, 2001, 45, pp 5-32

[35] Kotsiantis, S. B.: 'Decision Trees: a recent overview'. Artificial Intelligence Review, 2013, 39 pp 261-283

[36] http://www.stat.berkeley.edu/users/breiman/RandomForests/, Accessed October 2014

[37] Witten, H. I., Frank, E., and Hall, M. A.: 'Data Mining: Practical machine learning tools and techniques' Third edition, (Morgan Kaufmann, 2011)

[38] Landwher, N., Hall, M., and Frank, E.: 'Logistic model trees'. Machine Learning, 2005, 59(1-2), pp. 161-205

[39] Le Cisse, S., and van Houwelingen, J. C.: 'Ridge estimators in logistic regression'. Applied Statistics, 1992, 41 (1), pp. 191-201

[40] Cover, T. M., and Thomas, J. A.: 'Elements of Information Theory' (Wiley, 1991).

[41] Yerima, S. Y., Sezer, S., and McWilliams, G.: 'Analysis of bayesian classifcation-based approaches for Android malware detection'. IET Information Security, Vol 8, Issue 1, January 2014, pp 25-36